%% file: version2.tex
\documentclass[conference]{IEEEtran}
\IEEEoverridecommandlockouts
\usepackage{cite}
\usepackage{caption}
\usepackage{amsmath,amssymb,amsfonts}
\usepackage{algorithmic}
\usepackage{graphicx}
\usepackage{textcomp}
\usepackage{xcolor}
\usepackage{subcaption}
\usepackage[utf8]{inputenc}
\usepackage[T1]{fontenc} 
\usepackage{lipsum}
\usepackage{adjustbox}   
\usepackage{hyperref}
\usepackage{comment}
\usepackage{enumitem}
\usepackage{tikz}
\usepackage{enumitem}
\usepackage{soul}
\usepackage{booktabs}
\usepackage{tabularx}
\usetikzlibrary{er,positioning}
\usepackage{algorithmic}            
\usepackage{algorithm}
\usepackage{mathtools}
\DeclarePairedDelimiter\floor{\lfloor}{\rfloor}
\usepackage{adjustbox}

\def\BibTeX{{\rm B\kern-.05em{\sc i\kern-.025em b}\kern-.08em
    T\kern-.1667em\lower.7ex\hbox{E}\kern-.125emX}}
\input{math.tex}

\newcommand  {\newp}[1]{{\color{blue}\sf{#1}}}

\begin{document}

\title{Covariate Balancing Methods for Randomized Controlled Trials Are Not Adversarially Robust}

\author{\IEEEauthorblockN{Hossein~Babaei, Sina~Alemohammad, and
Richard~Baraniuk} 
\IEEEauthorblockA{
Rice University,
Houston, Texas, USA} 
{\{hb26, sa86, richb\}@rice.edu}

}

\maketitle

\begin{abstract}
The first step towards investigating the effectiveness of a treatment via a randomized trial is to split the population into control and treatment groups then compare the average response of the treatment group receiving the treatment to the control group receiving the placebo. 

In order to ensure that the difference between the two groups is caused only by the treatment, it is crucial that the control and the treatment groups have similar statistics. 
Indeed, the validity and reliability of a trial are determined by the similarity of two groups' statistics. Covariate balancing methods increase the similarity between the distributions of the two groups' covariates. However, often in practice, there are not enough samples to accurately estimate the groups' covariate distributions. In this paper, we empirically show that covariate balancing with the Standardized Means Difference (SMD) covariate balancing measure, as well as Pocock's sequential treatment assignment method, are susceptible to worst-case treatment assignments. 
Worst-case treatment assignments are those admitted by the covariate balance measure, but result in highest possible ATE estimation errors. 
We developed an adversarial attack to find adversarial treatment assignment for any given trial. Then, we provide an index to measure how close the given trial is to the worst-case.
To this end, we provide an optimization-based algorithm, namely Adversarial Treatment ASsignment in TREatment Effect Trials (ATASTREET), to find the adversarial treatment assignments.
\end{abstract}


\section{Introduction}

The standard method to measure the causal relationship between two variables is the Average Treatment Effect (ATE) \cite{Rubin1974}. 
The term ATE refers to the average outcome change that a certain intervention (which is called treatment) can make in a population in contrast to not making the intervention.

Randomized controlled trials (RCTs) are the gold standard for conducting quantitative experimental science \cite{chalmers1981method,hariton2018randomised ,benson2000comparison, concato2000randomized,deaton2018understanding}. 
RCT experimental design consists of recruiting a study population and
splitting the participants into two groups: treatment and control\footnote[1]{In this paper, we use terminology associated to medical clinical trials. However, any argument about medical clinical trials can be generalized to wider applications.}. If the treatment is assigned randomly, the difference between the
average outcomes of the two groups is an unbiased estimator of the ATE\cite{atheyreview}.
Since the trial is only conducted once, it is of high importance to reduce the ATE estimation variance.

Covariate Balancing Methods (CBMs) are methods to measure and induce more similarity in the statistics of the two groups. In order to ensure that the difference between the two groups is caused only by the treatment, it is crucial that the control and the treatment groups have similar statistics. The similarity of the statistics is commonly used to evaluate the validity and reliability of the conclusions based on the estimated ATE in an RCT. 

In this paper, we perform worst-case analysis of CBMs. We define the worst-case treatment assignments of a given CBM in an RCT as the treatment assignments that would be evaluated as sufficiently balanced by the given CBM, but would result in the highest possible ATE estimation error. We provide quantitative definition of {\em sufficiently balanced} later in the paper.

In this work, we perform worst-case analysis on two commonly used CBMs, the Standardized Means Difference(SMD) for non-sequential treatment assignments, and Pocock's sequential assignment method \cite{pocock1975sequential}. In both cases, we develop a method that finds the worst-case treatment assignments in a given RCT that we dub the Adversarial Treatment ASsignment in TREatment Effect Trials (ATASTREET). ATASTREET reduces the combinatorially large space of possible treatment assignments to efficiently find the worst-case treatment assignment.

In order to find worst-case treatment assignments, ATASTREET  works as an oracle method with the access to both potential outcomes. As an illustrative example, we use the  semi-synthetic IHDP1000 \cite{Gr,Hill2011,Sh} dataset, which provides both potential outcomes for each participant. IHDP is widely accepted as the standard benchmark dataset in heterogeneous treatment effect estimation. Naturally, some would criticize IHDP and argue that it is not a good reflection of a real-world RCT \cite{curth2021really}. Nevertheless, IHDP is still considered as the dataset that could provide the strongest evidence in heterogeneous treatment effect estimation literature.

We empirically demonstrate the worst-case vulnerability of the investigated CBMs. The worst-case treatment assignment can get selected for the trial as a result of CBM, either unluckily or by intentional deviations from a deceitful researcher. Since it results in maximally balanced groups, it encourages the confidence in the MATE with worst-case ATE estimation error. Since the trial is conducted only once, it is important to ensure that the selected treatment assignments is not close to worst-case treatment assignments.

We define \emph{CBM deviation index} $\rho$ to identify whether these worst-cases of CBM happened in any given RCT. This index provides a measurement on how close the selected treatment assignment is to the worst-case treatment assignments. For any given RCT, We use counterfactual estimation methods to estimate the unobserved potential outcomes. Then, ATASTREET finds the worst-case assignments. The $\rho$-index can be measured afterwards to identify the unlucky or deceitful deviations in the trial.

To further emphasis the importance of worst-case analysis and such sanity check, we develop an adversarial attack to any given RCT that used the mentioned CBMs, and empirically evaluate our introduced adversarial attack on the IHDP dataset. An adversary can exploit the adversarial vulnerability and use adversarial treatment assignments to maximize(or minimize) the measured ATE in the trial while having maximally balanced treatment groups.

We summarize our contributions as follows. First, we propose an optimization based algorithm (ATASTREET) to find worst-case treatment assignments of SMD and Pocock's method as CBMs. We then empirically demonstrate worst-case vulnerability of the mentioned CBMs.
Second, we provide an index to identify if a given trial is close to the worst-case assignment,
Third, we introduce an adversarial treatment assignment method using ATASTREET.
Finally, we demonstrate the adversarial vulnerability of SMD and Pocock's method and discussed some of the possible solutions to reduce the adversarial vulnerability.

\section{Background}
In this section, we first cover some of the basic definitions about ATE and RCTs, then discuss some recognized challenges. consequently, we cover how variance reduction techniques and covariate balancing methods are discussed in the literature.

\subsection{Background on Randomized Clinical Trails}

The ATE is defined using the potential outcome framework \cite{Rubin1974}. For each individual $i$ in the population, we call the potential outcomes of that individual being assigned to the treatment $Y_{i}^{1}$ or the control group $Y_{i}^{0}$. A set of covariates for each subject is also recorded as $\vec{x}^{\,i}$. The ATE is defined as the average of the differences of the potential outcomes for all the individuals over the population
\begin{equation}
    \text{ATE} =\; \dfrac{1}{N}\sum_{i} (Y_{i}^{1} - Y_{i}^{0}),
   \label{eq2}
\end{equation}
where $N$ is the population size.

In a trial to measure the ATE of a certain treatment(intervention), a {\em treatment assignment} $\mathcal{A} : \{ 1 , 2 , \dots , N \} \rightarrow \{ 0 , 1\}^{N}$ divides the population to either the treatment group or the control  group. For each individual, the $Y_{i}^{(\rm{obs})}$ is the the observed outcome based on the selected treatment assignment.
\begin{equation}
    Y_{i}^{(\rm{obs})} = \left\{
  \begin{array}{ll} 
      Y_{i}^{1} & \mathcal{A}(i) = 1 \\
      Y_{i}^{0} & \mathcal{A}(i) = 0 \;  .
      \end{array}
\right .
\label{eq3}
\end{equation}

The ``fundamental problem of causal inference''  \cite{Holland} is that each individual subject in the population can only be assigned to either the treatment or the control group. Therefore, the outcome of an individual subject given the treatment and that of the same individual not given the treatment cannot be observed in the same trial. As a result, half of the required data for estimating the ATE is unobservable.

Can this fundamental problem be solved? In \cite{dawid}, the author argued that estimating the unobserved potential outcomes can result in erroneous or metaphysical conclusions that are not substantiated by the data. Thus solutions for the ``fundamental problem of causal inference'' are dubious and cannot be supported by evidence in the experiment.  
\cite{Pearempicism} and \cite{shpitser} argued against this paradigm by providing a framework that, given some structural information about the causal relationships in the system, identifies cases where the unobserved potential outcomes can be discerned by observations.
Their arguments support the claim that the estimation of unobserved potential outcomes is a mathematical, not metaphysical, question. Some works first learn a causal graph over the variables with methods such as \cite{akbari2021recursive}; then study the ATE identifiability problem in the presence of unobserved variables. They argue that from the causal graph and observational data, some ATEs are non-identifiable due to the unmeasured confounders, and additional assumptions are required \cite{kivva2022revisiting,mokhtarian2022causal,shpitser2006identification}.

In the random treatment assignment method \cite{atheyreview}, The trial is conducted using a randomly selected treatment assignment $\mathcal{A} $. The Measured Average Treatment Effect (MATE) is then defined as 
\begin{equation}
    \text{MATE}( \mathcal{A} ) =\; \dfrac{1}{N_{1}}\sum_{\mathcal{A}(i)=1}  Y_{i}^{(\rm{obs})}  - \dfrac{1}{N_{0}}\sum_{\mathcal{A}(i)=0}  Y_{i}^{(\rm{obs})},
   \label{eq4}
\end{equation}
where $N_{0}$ and $N_{1}$ are the number of individuals assigned to the control and treatment groups, respectively.

Given the population, \cite{atheyreview} demonstrated that the introduced MATE is an unbiased estimator of the ATE. It means that the expected value of MATE over the random treatment assignment $\mathcal{A}$ is equal to the true value of ATE.

 \emph{The ATE estimation error} for any given treatment assignment $\mathcal{A}$ is the error in the MATE when $\mathcal{A}$ is used as the treatment assignment
\begin{align}
    \Tilde{\epsilon} (\mathcal{A}) = \rm{MATE}(A)- \rm{ATE} \; .
   \label{eq_mate_error}
\end{align}

Generally, the goal is to reduce $|\Tilde{\epsilon} (\mathcal{A})|$ as much as possible.

\subsection{Challenges in Randomized Clinical Trials}

The estimated ATE has some variance due to randomly selected treatment assignment. The mentioned \emph{ATE variance} is the variance of the ATE estimation when $\mathcal{A}$ is selected uniformly random
\begin{align}
    \sigma^2_{\rm{ATE}}  = \E_{\mathcal{A}\in \{ 0,1\}^N  }\left[ \Tilde{\epsilon}^2   \right ]. 
   \label{eq_mate_var}
\end{align}

\begin{figure}[htbp]
\centerline{\includegraphics[scale=0.4]{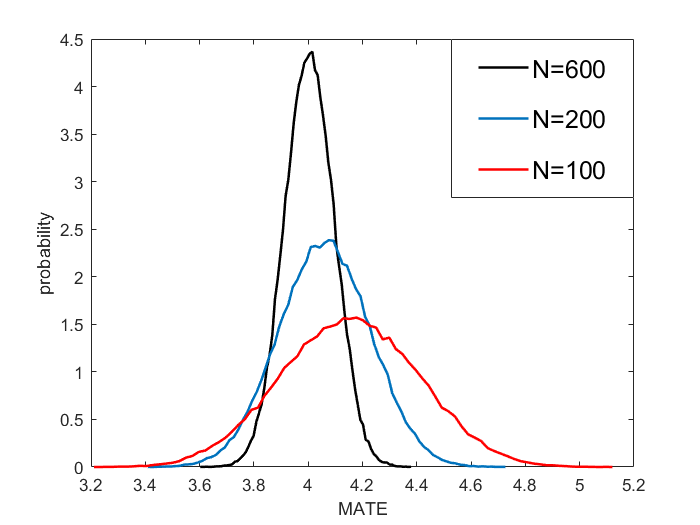}}
\caption{The empirical probability distribution of the MATE in IHDP1000 for varying population sizes $N$. As the population size grows, the variance $\sigma^2_{\rm{ATE}}$ decreases.}
\label{fig1}
\end{figure}

Although the MATE estimator is unbiased, it is a single observation estimate since the trial is typically conducted only once. As a result, there is uncertainty in the MATE. Another possible way to control the variance in MATE is to increase the population size used in the RCT. However, the variance can still be undesirably large for the affordable population size. In order to empirically show this issue, we measured the MATE for 10000 different random treatment assignments in the IHDP dataset \cite{Gr,Hill2011,Sh} for different sub-population sizes. Figure~\ref{fig1} shows the empirical probability density distribution of the MATE. Clearly, the variance shrinks as the population size grows; however, variance might still be undesirable in sensitive tasks for the affordable population sizes (in this case $N<600$). 

Since a typical trial is conducted once, only one treatment assignment can be used for the trial. Thus, it is of high importance to ensure that one selected treatment assignment is selected properly\cite{atheyreview}. Even in the case of proper randomization, it may be important to check whether the selected treatment assignment has imbalanced covariates by chance. Furthermore, it is common in practice that some participants drop-out before the trial is finished. The drop-outs make the trial population different from the original population which was used in the randomization, which in turn might induce a selection bias. For all of the mentioned reasons, it is important to check for baseline imbalances.

Sometimes  $p$-value based hypothesis testing is used in order to check whether the treatment assignment is selected properly. This usage has been recognized as illogical \cite{R11senn1994testing,R13harvey2010electrical,R14altman1985comparability}. “Such significance tests assess the probability [i.e., P-value] that observed baseline differences could have occurred by chance; however, we already know that any differences are caused by chance.” \cite{R121schulz2010consort,R122schulz2010consort}  
As a result, $p$-values for baseline differences does not serve a useful purpose since it isn’t testing a useful scientific hypothesis\cite{R115austin2010substantial,R11senn1994testing,R14altman1985comparability,R2pocock2002subgroup}. Later in this section balancing scores are discussed as tools that should be used to evaluate the baseline comparability. 

\subsection{MATE Variance Reduction}

There have been numerous efforts to reduce the
estimation variance of the MATE. Covariate adjustment and CBMs are two families of such efforts.

Covariate adjustment tools reduce the effects of baseline imbalances on the estimated ATE by using different regression models.

Some believe that any dissimilarity in the statistics of the two groups can be compensated using covariate adjustment methods, such as ANCOVA \cite{keselman1998statistical,van2006ancova,rutherford2011anova,wright2006comparing, johnson2016violation, jamieson2004analysis}. Thus it is of no interest to test for similarity of statistics in the two groups, or try to use treatment assignments with more similar statistics\cite{senn1994testing}.

Several authors have argued against this belief in four main arguments:

i) Covariate adjustment tools have complex statistical properties. Thus, unadjusted findings are preferred by authors and readers because such findings are simpler and have more clarity \cite{R2pocock2002subgroup}. It explains why even when deployed, covariate adjusted findings are mostly used as the backup for the unadjusted findings \cite{R2pocock2002subgroup}.

ii) It has been shown that different models can lead to various estimates and maybe even different clinical implications. Potential biased choices out of numerous different model families and parameter settings is one of the reasons of suspicion regarding the potential manipulations of covariate adjustment methods which ultimately make them less credible \cite{R2pocock2002subgroup}.

iii) In some trials, covariate adjustment methods need more than affordable population size in order to adjust for all the covariates. As a result, those covariates that are expected to be more prognostic would be adjusted. In some trials, there is insufficient clinical agreement or there's lack of confidence on which covariates should be adjusted for \cite{R2pocock2002subgroup}.

iv) \cite{pearl2009causality,huang2006identifiability,mokhtarian2022causal} have also studied the ATE identifiability problem and argued that in some cases, it is not possible to identify ATE in the presence of biases as they introduces some unmeasured confounders to the underlying causal graph.

 Authors in \cite{R2pocock2002subgroup} have summarized these arguments as:" The scope for judgements in an ill-defined strategy, and biased (for example, most favorable) choices out of a multiplicity of possible analyses, means that covariate adjusted analyses may rightly be viewed with some suspicion, often leaving primary emphasis on the unadjusted analysis"

A common practice in trial reports is to devote "Table I" (also known as patient cohort) to comparing the distributions of baseline variables among different treatment groups. In addition to the fact that it helps the reader to decide whether this study can be generalized to another population, there are two main goals in having separate columns for different treatment groups rather than just a single column for the whole population. First, it demonstrates that the randomization worked well, or it can identify any unlucky imbalances. Second, having balanced baseline variables adds credibility to the trial, especially encouraging confidence in the unadjusted analysis \cite{R2pocock2002subgroup}.

Can covariate adjustment substitute the need for baseline comparability? Although covariate adjustment tools have numerous benefits, following previous paragraphs, they cannot substitute the need for baseline comparability and balanced covariates.

\subsection{Covariate Balancing Methods}

CBMs are a family of methods in which treatment assignments with more similarity in the statistics of two groups have a higher chance to be selected for the trial. In CBMs, all of the variables that are expected to be related to the outcome are recorded for the population as the covariates. CBMs try to favor treatment assignments that have more similarity between the covariates' distributions in the two groups. Since the treatment and the control group are ``similar" in such balanced treatment assignments, selection bias can thereby be reduced.

CBMs require a balancing score (also referred to as the covariate balance measure) that evaluates the similarity of the covariate distributions of the control and treatment groups.

The common motivation behind all of the CBMs is to promote similarity of the joint distribution of covariates between the two groups. In the mathematical language, if covariates of each subject are recorded as $\vec{x}^{\,i}$ then $P(\vec{x})$ for the treatment and the control groups should be similar. With the limited population size and high number of covariates, promoting and measuring this similarity becomes intractable in practice. That's where different CBMs relax the problem in different ways.

There are two main categories of RCTs, the non-sequential RCTs where covariates of the whole population are assumed to be accessible before the conductance of the trial, and the sequential RCTs where subjects become available sequentially. Sequential and non-sequential CBMs are targeted towards the sequential and non-sequential RCTs respectively.

\subsubsection{Non-sequential CBMs}

The first step of non-sequential CBM in RCTs includes recording the covariates for the population. Then, balanced treatment assignments are found by minimizing the covariate imbalance among the two groups. In the next stage, the trial is conducted according to the obtained balanced treatment assignment. The MATE, then, is calculated afterwards.

There are different implementations for a given CBM. An initial treatment assignment can be selected randomly and then a greedy minimization modifies the treatment assignment until it reaches a desirable balancing score\cite{Thereview}. Alternatively, the whole randomization process can be repeated until a treatment assignment with a desired balance is reached \cite{Thereview}. Another option is that one exhaustively checks all possible treatment assignments in order to find the treatment assignment that is maximally balanced. Alternatively, one can find a set of acceptable treatment assignments, and then select one of them randomly.

One of the most commonly used CBMs is Standardized Means Difference (SMD), the difference of the means of each covariate between the treatment and the control group. In order to avoid scaling issues, this CBM standardizes the difference of the means of each covariate by the variance of that covariate \cite{Thereview,rosenbaum1985b}.

The balancing score for SMD is defined as
\begin{align}
    \mathcal{U}_{p} =\; \left\|  
    \dfrac{1}{N_{1}} \sum_{ \rm{treatment}}\vec{x}^{\,i} - \dfrac{1}{N_{0}} \sum_{\rm{control}} \vec{x}^{\,i} 
    \right\|_{p}, \, p \in \{1, \infty\},
   \label{eq_pnorm}
\end{align}
where $N_1$ and $N_0$ are the size of the treatment and control group, respectively. And $\vec{x}^{\,i}$ is a vector containing the covariates of the $i^{\textit{th}}$ subject. Both $\ell_{1}$ and $\ell_{\infty}$ can be used for vector norms in cases with more than one covariate.

We assume that all of the covariates have the same variance without loss of generality. If that is not the case, one can simply normalize each covariate by its standard deviation.

Some other non-sequential CBMs has also been proposed. In \cite{rubin2001} three different CBMs are proposed based on the propensity score as a scalar representation for the covariates of each individual. Using the propensity score concept, the three proposed CBMs are 1) the difference of means of the propensity scores normalized to the variances, 2) the ratio of the variance of the propensity scores in the control and the treatment group, and finally, 3) the the ratio of the variance of each covariate orthogonal to the propensity score in the treatment and the control group.

\subsubsection{Sequential CBMs}

Another recognized category of CBMs is sequential treatment assignment. In many of the trials, especially in the medical trials, the whole population is not accessible at once, and the population recruitment is performed sequentially. Even if the whole population is available at the beginning of the trial, there is always a possibility that some of them drop out from the trial or more subjects get added to the trial in order to increase quality of the results. The sequential treatment assignment can handle the mentioned situations.

One of the most popular sequential treatment assignment methods is proposed by Pocock \cite{pocock1975sequential}. We highly encourage the reader to study this method from the original source but we include a simplified executive summary of its binary version as Algorithm~\ref{pcok} in the appendix for the ease of convenience. 
 
Several other sequential treatment assignment methods have also been proposed to promote covariate balance \cite{atan2018adaptive ,pocock1975sequential ,taves1974minimization}.


In this paper, we investigate worst-case vulnerability of one of the most used CBMs in each category of sequential and non-sequential treatment assignment. SMD is one of the most used non-sequential CBMs \cite{atheyreview , Thereview,R1nguyen2021incomparability}
\cite{R3imai2008misunderstandings,R4austin2009balance,R5ali2015reporting,R6austin2007propensity,R7austin2008critical,R8gayat2010propensity,R9lonjon2017potential,R10cochran1968effectiveness} We also investigate Pocock's sequential assignment method as one of the well-known sequential CBMs.
 
SMD compares the means of the two joint distributions and forces covariates in different groups to have similar means. On the other hand, Pocock's sequential treatment assignment method promotes similarity in the marginal distributions of different covariates; which is a stronger similarity than the SMD. In the next sections, we provide arguments on the effects of promoting stronger similarity on the adversarial vulnerability.

\section{Worst-Case Treatment Assignments}

In this paper,  for the first time, we empirically find worst-case treatment assignments for the SMD and Pocock's sequential assignment method. Then we analyse the empirical results in order to study worst-case behaviors of the given CBMs.

\subsection{Definitions}

In order to formally define worst-case treatment assignments, some concepts should be defined beforehand;

\emph{The covariate balancing score} $\; \mathcal{U}\; $ (also referred to as the balancing measure) is a scalar function that returns the amount of covariate imbalance of a given treatment assignment. Note that a higher covariate balancing score means that the treatment assignment is more imbalanced.
\emph{The expected imbalance} $\; \bar{\mathcal{U}}\; $ is the expected value of the covariate balance measurement $\mathcal{U}$ over all the possible treatment assignments in the trial. 
\emph{The minimum imbalance} $\; \mathcal{U}_{min}\; $ is the minimum value of $\mathcal{U}$ over all the possible treatment assignments in the trial.

\emph{The admissible treatment assignment set} $ \Tilde{\mathcal{A}}$   is defined as the set of all the treatment assignments 
\begin{align}
    \Tilde{\mathcal{A}} = \left\{ \mathcal{A} \; | \forall{ \mathcal{A}' } \; , \; \frac{\mathcal{U}(\mathcal{A}) - \mathcal{U}(\mathcal{A}')}{\bar{\mathcal{U}}} < \alpha_a  \right\},
   \label{eqadissible}
\end{align}
where $\alpha_a \ll 1$ is a parameter that controls the amount of balance induced by the CBM. Larger $\alpha_a$ relaxes the covariate balancing and allows for more treatment assignments to be admissible.

In order to quantify the vulnerability of a given RCT to worst-case treatment assignments, we measure the maximum possible deviation of MATE in the admissible treatment assignments set.

We define \emph{worst-case deviation factor $\xi$} as the range of the measured ATE by different admissible treatment assignments, normalized by the standard deviation of the measured ATE over random treatment assignments
\begin{align}
    \xi &= \dfrac{\rm{Range} [\rm{MATE}( \Tilde{\mathcal{A}})] }{2\sigma_{\rm{ATE}}} .
   \label{eqxi}
\end{align}

\subsection{Worst-case assignments for SMD in Non-Sequential Trials}

We are interested in finding worst-case treatment assignments of the SMD  as CBM in the trial.

{\bf{Worst-case treatment assignment for SMD: }} Assume that the potential outcomes of assigning each subject to the treatment or the control group are provided for a population size of $N$. The potential outcome for the subject $i$ being assigned to the treatment group or the control group is $ y^{1}_{i} $  and $ y^{0}_{i} $, respectively. For each subject in the population, covariates are provided as an $M$-dimensional vector $\vec{x}^{\,i}$ . The goal is to find the treatment assignment $ \mathcal{A}^{*} : \{ 1 , 2 , \dots , N \} \rightarrow \{ 0 , 1\}^{N}$ dividing the population into two groups with equal sizes such that it maximizes the MATE and minimizes the covariate balancing score $\mathcal{U}_p$. We use Lagrange multipliers to formulate a combinatorial optimization problem over the space of all possible treatment assignments
\begin{align}
    \mathcal{A}^{*} _{p}= \mathrm{argmax}_{\mathcal{A}}  \Bigg( \lambda\, \mathrm{MATE}(  \mathcal{A} ) -  \mathcal{U}_{p} ( \mathcal{A} )  \Bigg), \, p =\{ 1,\infty\}. 
   \label{eq6}
\end{align}

The above problem is a combinatorial optimization problem over the space of all possible treatment assignments. ATASTREET converts the above problem to a constrained linear programming problem and solves it using mixed integer linear programming tools in an acceptable time \cite{milp,KLOTZ,Narendra,NAU,Clausen,Land,Bader2005}. More details are provided in the appendix.

\subsection{Worst-Case Treatment Assignments for Sequential Trials}

Finding worst-case treatment assignments of the sequential CBMs is even more challenging since the treatment assignment of one subject affects the treatment assignment of the next subjects. We approach this challenge by providing a non-sequential balancing score
\begin{align}
    \mathcal{U}_{\rm{Pocock}} =\sum_{i=1}^{m} \sum_{j=1}^{N_i} \alpha_i |N_{\rm{control}}^{i}(j) - N_{\rm{treatment}}^{i}(j) |,
   \label{eq_Pocock_def}
\end{align}
where $m$ is the number of covariates, $N_i$ is the total number of categories for $i^{\textit{th}}$ covariate,\footnote{Recall that the covariates are assumed to be categorical in Pocock's sequential assignment method in Algorithm~\ref{pcok}.} and $N_{\rm{control}}^{i}(j)$
is the total number of subjects in the control group with their $i^{\textit{th}}$ covariate having the value of $j^{\textit{th}}$ category, and $N_{\rm{treatment}}^{i}(j)$ is the same for the treatment group. Then, we provide a theorem that tightly links our proposed balancing score to Pocock's sequential treatment assignment method (Algorithm \ref{pcok}).

\begin{thm}
In Pocock's sequential treatment assignment method,  using $\mathcal{U}_{\rm{Pocock}}$ instead of $G$ in Pocock's method (Algorithm~\ref{pcok}) results in the same decision rule.
\end{thm}

For the proof, see the appendix.

The above theorem suggests that Pocock's sequential method is in fact a sequential greedy probabilistic minimization over a non-sequential CBM with $\mathcal{U}_{\rm{Pocock}}$ as its balancing score. Putting the randomnesses aside, Pocock's sequential method favors treatment assignments with smaller $\mathcal{U}_{\rm{Pocock}}$. The goal of our worst-case analysis of Pocock's method would be to find treatment assignments that are favored by Pocock's method the most, and have maximum possible ATE estimation error.

 Arguments in the previous paragraph motivate us to find the adversarial treatment assignments of the mentioned non-sequential CBM. Then, each of the resulting worst-case treatment assignments should carefully be analysed to see whether they are feasible to get selected by Pocock's sequential method.

{\bf{Worst-case treatment assignment for Pocock's CBM: }} Assume that $ y^{1}_{i} $ , $ y^{0}_{i} $, and $\vec{x}^{\,i}$ are given similar to worst-case analysis for SMD. The goal is to find the treatment assignment $ \mathcal{A}^{*} : \{ 1 , 2 , \dots , N \} \rightarrow \{ 0 , 1\}^{N}$ dividing the population into two groups with equal sizes such that it maximizes the MATE and minimizes the covariate balancing score $\mathcal{U}_{\rm{Pocock}}$. We use Lagrange multipliers to formulate a combinatorial optimization problem over the space of all possible treatment assignments
\begin{align}
    \mathcal{A}^{*} _{\rm{Pocock}}= \mathrm{argmax}_{\mathcal{A}} \;  \Bigg( \lambda \; \mathrm{MATE}( \mathcal{A} ) -  \mathcal{U}_{\rm{Pocock}} ( \mathcal{A} )  \Bigg).
   \label{eq_atastreet_pocock}
\end{align}

Similar to the previous case where we covered SMD for non-sequential RCTs, we obtain ATASTREET solution using mixed linear integer
programming \cite{milp,KLOTZ,Narendra,NAU,Clausen,Land,Bader2005}. More details are provided in the appendix.

\begin{figure*}[t]
\begin{subfigure}{.33\textwidth}
  \centering
  \includegraphics[width=\linewidth]{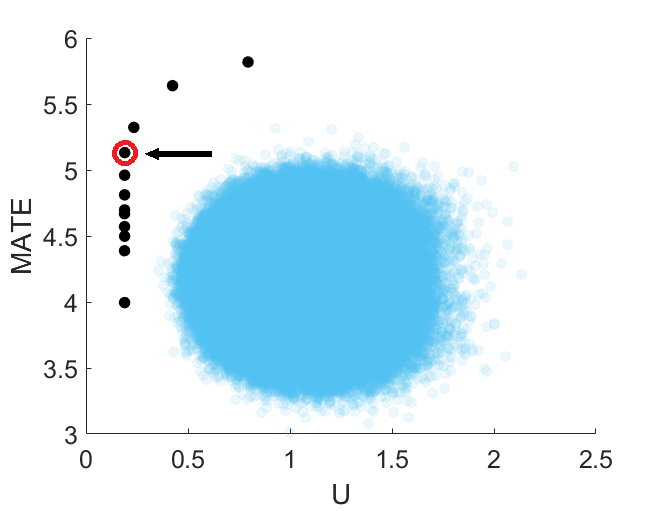}  
  \caption{SMD with $\mathcal{U}_{1}$}
\end{subfigure}
\begin{subfigure}{.33\textwidth}
  \centering
  \includegraphics[width=\linewidth]{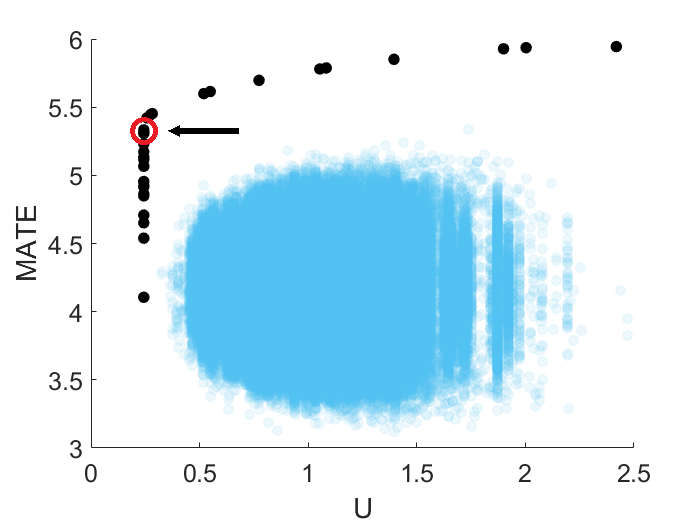}  
  \caption{SMD with $\mathcal{U}_{\infty}$}
\end{subfigure}
\begin{subfigure}{.33\textwidth}
  \centering
  \includegraphics[width=\linewidth]{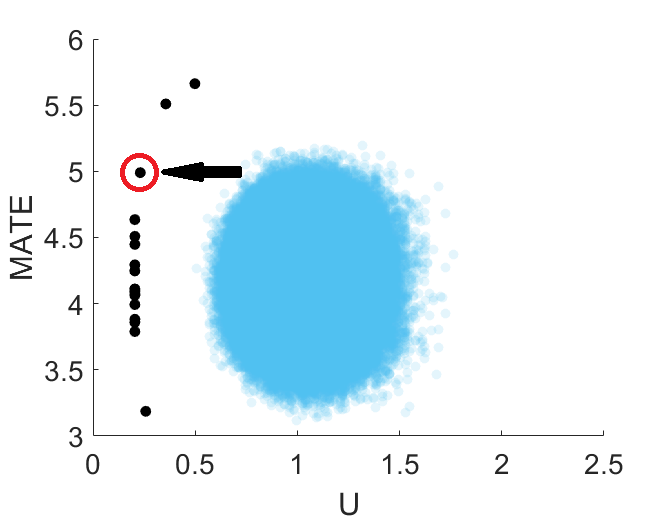}  
   \caption{$\mathcal{U}_{\rm{Pocock}}$}
\end{subfigure}
\caption{ We visualize ATASTREET results for different $\lambda$s on the IHDP1000 dataset (black points). A reference set of randomly selected treatment assignments of IHDP is also visualized as blue dots. The $\mathcal{U}_p$ axis is normalized to $\bar{\mathcal{U}}$. This plot shows ATASTREET solutions in comparison to the reference set (blue points). The marked red circled points could have been selected in the IHDP trial. The perfect balance of covariates would encourage confidence in the trial with a large ATE estimation error. This plot shows vulnerability of the mentioned CBMs to worst-case treatment assignment.}
\label{fig2}
\end{figure*}

\subsection{Empirical Results of Worst-Case Analysis}

In order to provide a better understanding of worst case treatment assignments, we introduce a new visualization technique for different possible treatment assignments in the same trials.
Each treatment assignment is visualized as a single point with its corresponding $\mathcal{U}$ as the horizontal coordinate, and its corresponding MATE as the vertical coordinate.

We used our introduced visualization technique in order to visualize ATASTREET's resulting treatment assignments for different parameter $\lambda$ (Shown as black point in Figure~\ref{fig2}). A set of random treatment assignments with no CBM is also shown in each plot with blue points to act as a reference.


Several remarks follow from these results.

The CBMs in both our cases, the SMD for non-sequential case and Pocockc's method for sequential case, are vulnerable against worst-case treatment assignments. Analyzing the ATASTREET's resulting treatment assignments for different values of $\lambda$ reveals some of the worst-case treatment assignments(See Figure~\ref{fig2}). According to the results of our experiments, $\xi>6$. In another language, it is possible to find admissible treatment assignments where groups are well-balanced, but the MATE has error higher than $6\sigma_{\rm{ATE}}$. 

Following the previous argument, both analyzed CBMs are vulnerable against worst-case assignments. This vulnerability opens up unwanted potentials for deviations (intended or unintended) with considerable effects on the MATE. Restricting such potentials is very important in some applications like medical trials. In Figure~\ref{fig2},  the corresponding treatment assignment of the black point marked with the red circle is admissible with regards to having balanced covariates, yet yields a larger ATE estimation error than all the $10^6$ random treatment assignment shown as blue points .

\begin{figure}[t]
    \centering
    \includegraphics[width=.8\linewidth]{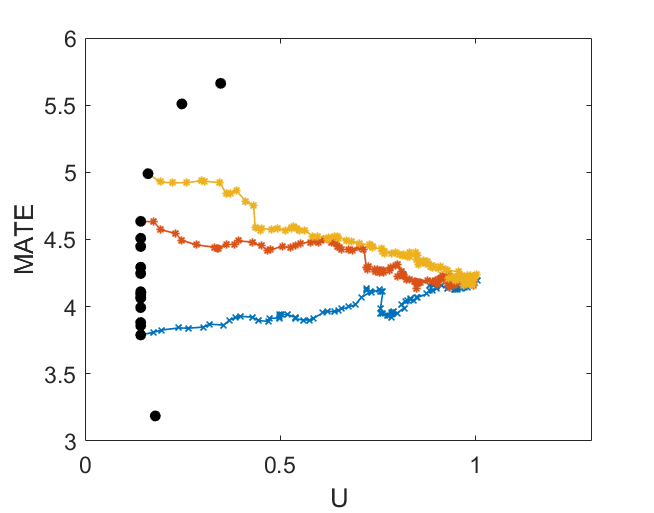}  
     \caption{
     Evolution of the treatment assignment sequence as new subjects are introduced to the trial. The $\mathcal{U}$ axis here represents the expected value of final $\mathcal{U}$ if the next subjects are to be assigned randomly (the expected value is approximated using Monte-Carlo method). This plot shows how Pocock's sequential assignment method reaches the worst-case treatment assignments found using ATASTREET (Equation~\ref{eq_atastreet_pocock}). 
     }
    \label{figfeasibility}
    \end{figure}

In the sequential case, it's not clear whether the worst-case treatment assignments associated with $\mathcal{U}_{\rm{Pocock}}$ are feasible to get selected by Pocock's sequential method. To demonstrate their feasibility, we considered different orders of subjects coming into the trial, and we set $P_0=1$ (Algorithm~\ref{pcok}) to ensure that Pocock's sequential method would never go towards the unlikely path. We found several different subject ordering where the evolution path goes to any of the predetermined assignments in ATASTREET results Equation~\ref{eq_atastreet_pocock}.  Although we do not provide any theoretical proof that ATASTREET solutions are always feasible for selection by Pocock's method with $P_0=1$, we have empirically provided several different paths for each of the resulting ATASTREET's assignments (Figure~\ref{figfeasibility}).
Furthermore, oftentimes, $P_0<1$ in practice. It means that any treatment assignment is now possible to get selected by Pocock's sequential method. Arguments regarding posterior probability of worst-case assignments getting selected is out of the scope of this paper. To summarize arguments in this section, we have empirically found treatment assignment evolution paths that Pocock's sequential assignment method ends up in each of the worst-case assignments (Figure~\ref{figfeasibility}).
    
Our empirical results for different choice of $\mathcal{U}_{1}$ and $\mathcal{U}_{\infty}$ as different versions of SMD suggests that our arguments do not depend on the vector norm used in the SMD (Equation~\ref{eq_pnorm}). We infer that the observed vulnerability is inherent in the SMD, and not the deployed vector norm.

Pocock's method is slightly better than SMD(Figure\ref{figzetas}). Even though Pocock's method has smaller worst-case deviation factor $\xi$, it's still vulnerable and more CBMs should be investigated to find CBMs with smaller $\xi$s. One can modify ATASTREET for different CBMs to find their worst-case treatment assignments and compare their worst-case deviation factors $\xi$, Ultimately, the most worst-case robust CBM could be identified. Such CBM is ideal in cases where the clinical implications of the RCT is important and large errors in ATE estimation would inflict intolerable losses to health or financial resources.


\section{How Close is a Trial to Worse-Case?}

In the previous section, we have empirically demonstrated that the two investigated CBMs are vulnerable to worst-case assignments. It brings up an important question. \emph{How to ensure a trial is not close to the worst-case?} We answer this question by providing the \emph{CBM deviation index $\rho$}. 

A variety of ITE estimation tools can be used in order to assess the estimated ATE error for a given RCT \cite{yoon2018ganite,alaa2019validating,qian2021synctwin,curth2021nonparametric,johansson2016learning,crump2008nonparametric,chipman2010bart,alaa2017bayesian,breiman2001random, wager2018estimation }. Once the error interval is acquired, one can simply compare it to $\sigma_{\rm{ATE}}$ to interpret it as a unit-less number. In that case, the deployed treatment assignment is compared to random treatment assignments without CBM. In order to interpret the ATE estimation error in RCTs where CBM is used, we suggest comparing the ATE estimation error to the worst-case error in the similar balancing scores.

We define the \emph{CBM deviation index $\rho$} as the ratio of ATE estimation error to the worst-case error in the similar balancing score. The actionable summary on how to measure $\rho$ in any given trial without having access to unobserved counterfactual outcome is provided as Algorithm~\ref{CBM_deviation_index_alg} 

\begin{algorithm}[t]
\caption{CBM deviation index $\rho$ }
\label{CBM_deviation_index_alg}
    \begin{algorithmic}[1]
    \STATE Using the state-of-the-art ITE estimation method, estimate the unobserved counterfactual outcomes for all the subjects.
    
    \STATE Approximate the ATE estimation error $|\tilde{\epsilon}(\mathcal{A})|$ in the RCT using the estimated potential outcomes.
    
    \STATE Run ATASTREET on the resulting RCT table. Use a set of different parameter $\lambda$ In ATASTREET.
    
    \STATE Make a plot similar to Figure~\ref{fig2}.
    
    \STATE Connect the resulting ATASTREET treatment assignments so that they form a continuous contour. For this purpose, the finer sweep for parameter $\lambda$ results in a better approximation of the mentioned contour.
    
    \STATE Using the deployed treatment assignment in the RCT, calculate the balancing score.
    
    \STATE In ATASTREET contour, find a point with a balancing score equal to the balancing score of the deployed treatment assignment in the trial. This $\tilde{\epsilon}_{\rm{max}}$ is the maximum possible ATE estimation error in treatment assignments with similar balancing scores.
    
    \STATE Report the CBM deviation index $\rho=\dfrac{|\tilde{\epsilon}(\mathcal{A})|}{\tilde{\epsilon}_{\rm{max}}}$.

    \end{algorithmic}
\end{algorithm}

ITE estimation methods provide a noisy imperfect estimate of the ITE as well as the unobserved potential outcomes for each subject. Using these methods to estimate the unobserved counterfactual outcomes compromises the efficiency of worst-case assignments found by ATASTREET.

To investigate this, we formed a reconstructed version of IHDP by picking a realization of IHDP, then picked a treatment assignment at random and gave only the observed outcomes and deployed treatment assignment to GANITE \cite{yoon2018ganite}. Then , the estimated unobserved potential outcomes and the observed outcomes would form our reconstructed version of IHDP.

To investigate the effect of using noisy estimates of unobserved potential outcomes, we took 15 random realizations of the reconstructed version of IHDP and found worst-case treatment assignments using ATASTREET. Then we used ground truth from IHDP and measured the ground truth for ATE of the resulting assignments, Our results suggest that this imperfection resulted in estimating the \emph{worst-case deviation factor $\xi$} as $5$ times smaller than it's true value. Indeed, using better ITE estimators results in better measurements of the worst-case deviation factor $\xi$ as well as the CBM deviation index $\rho$.

\section{Towards Adversarial Attacks \\ of Clinical Trials}

In this section, we develop an adversarial attack to RCTs with mentioned CBMs. To do this, we provide an actionable summary of how to find adversarial treatment assignments for any given trial using ATASTREET.

In the previous sections, we empirically demonstrated that the mentioned CBMs are vulnerable to worst-case treatment assignments. We then provided an index to check whether a given RCT is close to the worst-case. In order to further emphasise the importance of such sanity check we develop an adversarial attack to any given RCT and demonstrate that an adversary can use such attack in order to deceitfully deviate the MATE while having maximally balanced groups.

Can someone exploit this vulnerability and find adversarial treatment assignment in a given RCT? We uncovered the worst-case assignments of the given CBMs using ATASTREET as an oracle method which has access to the ground truth values of the unobserved counterfactual outcomes. In this section, we provide an actionable summary on how to find adversarial treatment assignments in any given RCT.

For any given RCT, pick the state-of-the-art ITE estimation method, and use the observed outcomes as well as the deployed treatment assignment to estimate the unobserved counterfactual outcomes for all the subjects, then form the reconstructed version of the given trial. We argue that the worst-case assignments of the reconstructed version serve as adversarial assignments for the given RCT.

To empirically demonstrate this argument, we took 15 random realizations of IHDP1000, then formed the reconstructed version similar to the previous section by  removing  half of the observed potential outcomes and estimating them using GANITE. We found worst-case assignments of the reconstructed version, and used the ground truth values of potential outcomes in IHDP1000 in order to evaluate the resulting MATE of the adversarial treatment assignments. In Table \ref{table:1} and \ref{table:2}, $\tilde{\epsilon}_{\mathrm{adv}}$ is the resulting ATE estimation error of our adversarial attack normalized to $\sigma_{\rm{ATE}}$, $\xi$ is the worst-case deviation factor in IHDP, and $\rho$ is the efficiency of our introduced attack. As our result suggest, our introduced adversarial attack results in $\rho=0.2$, which means that our introduced adversarial attack has the ATE estimation error $5$ times smaller than the worst-case assignment.

Using ITE estimators with better accuracy results in less estimation error in reconstruction of the RCTs. Counterfactual outcome estimation and ITE estimation are active research areas and introducing methods with higher accuracy, results in adversarial treatment assignments closer to the worst-case assignments(bigger $\rho$).

\begin{table}
\label{Table_pocock}
\centering
\caption{Pocock's method}
{
\begin{adjustbox}{width=0.70\linewidth,center}
\begin{tabular}{|l|ccc|}
\toprule
\textbf{}         & \textbf{$\tilde{\epsilon}_{\mathrm{adv}}$} & \textbf{$\xi$} &\textbf{$\rho$} \\ \hline
Mean               & 1.50       & 7.87      & 0.20      \\
Std                 & 1.25       & 6.34     & 0.08 \\
Max                    & 5.33       & 24.05    & 0.37       \\
\bottomrule
\end{tabular}
\label{table:1}
\end{adjustbox}
}
\end{table}

\begin{table}
\centering
\caption{SMD with $\ell_{\infty}$}
{
\begin{adjustbox}{width=0.70\linewidth,center}
\begin{tabular}{|l|ccc|}
\toprule
\textbf{}      &   \textbf{$\tilde{\epsilon}_{\mathrm{adv}}$} & \textbf{$\xi$} &\textbf{$\rho$} \\ \hline
Mean               & 1.51       & 5.93      & 0.21      \\
Std                 & 1.63       & 2.33     & 0.13 \\
Max                    & 6.25      & 11.89     & 0.52       \\
\bottomrule
\end{tabular}
\label{table:2}
\end{adjustbox}
}
\label{tab:real-data-svm}
\end{table}

\begin{figure}[t]
\begin{subfigure}{.23\textwidth}
  \centering
  \includegraphics[width=\linewidth]{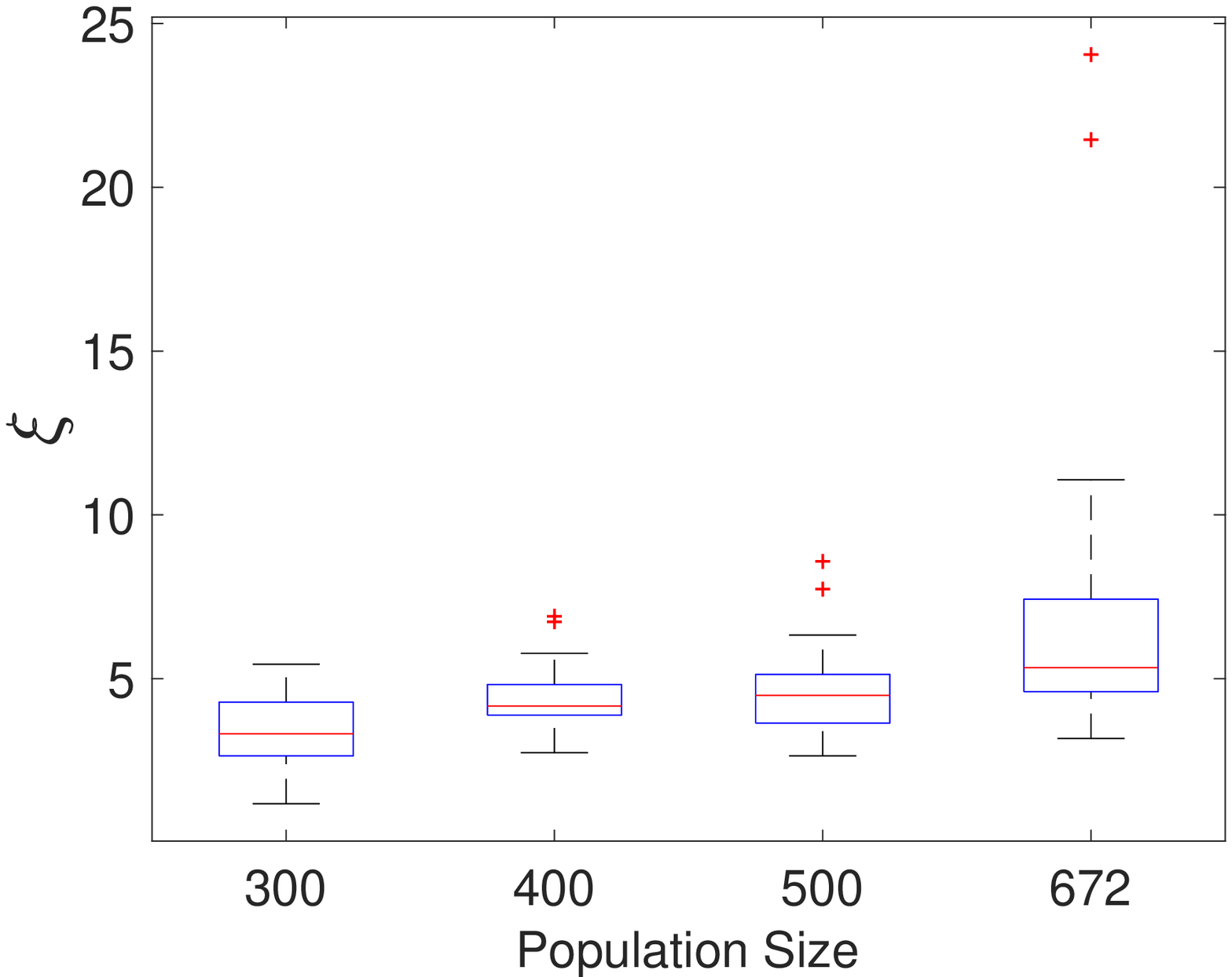}  
  \caption{Pocock's method}
\end{subfigure}
\begin{subfigure}{.23\textwidth}
  \centering
  \includegraphics[width=\linewidth]{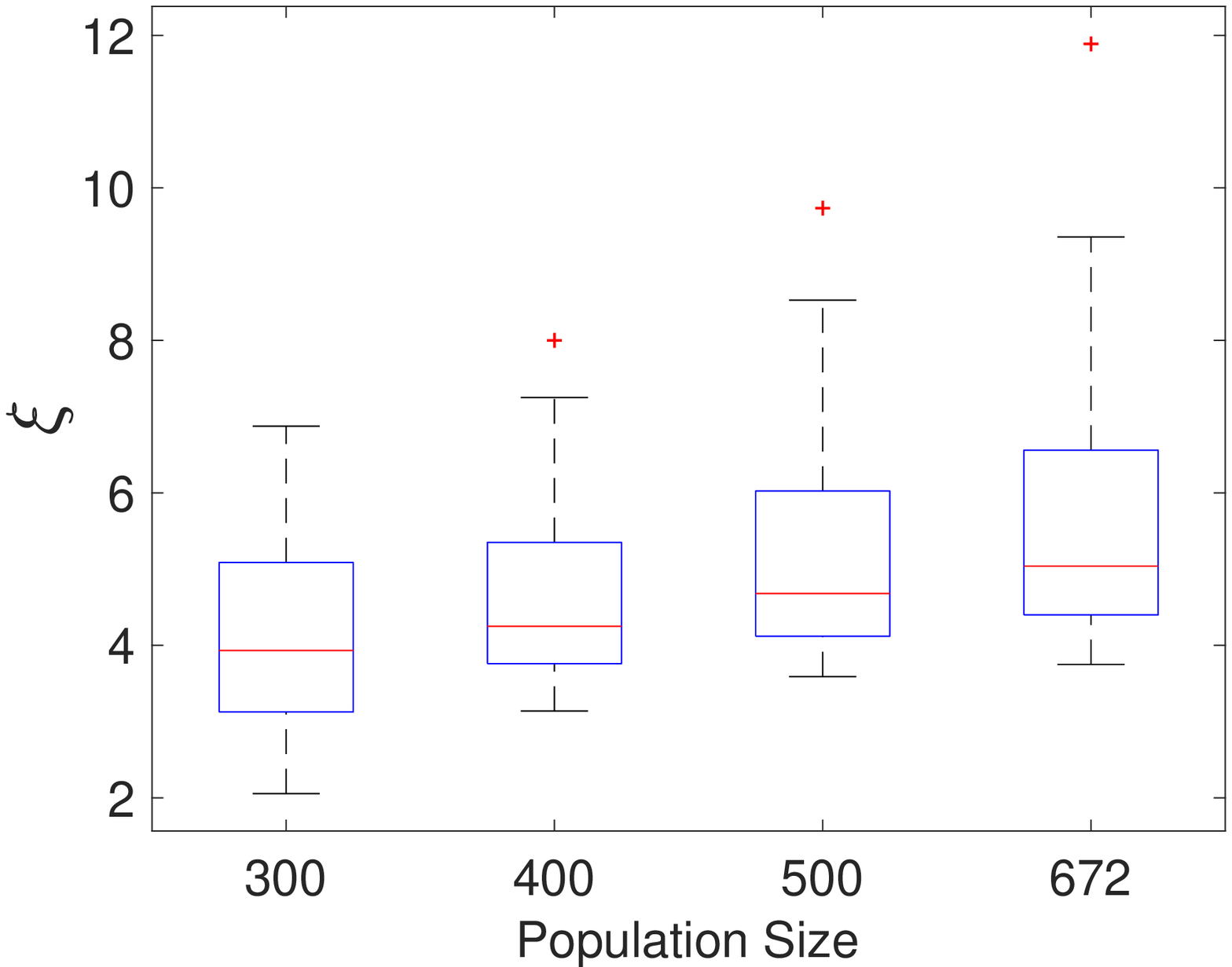}  
  \caption{SMD with $\mathcal{U}_{\infty}$}
\end{subfigure}
\caption{ worst-case deviation factor $\xi$ for different population sizes measured for $\alpha_a=0.02$ (Equation~\ref{eqadissible},\ref{eqxi}).} \label{figzetas}
\end{figure}

We investigated the effect of population size on the adversarial vulnerability of the analyzed CBMs. To do this, we randomly sub-sampled a population from the original population and found ATASTREET solutions, then plotted the resulting adversarial vulnerability factor $\xi$ for different population sizes in Figure~\ref{figzetas}. Unlike the MATE variance that shrinks with increasing the population size, the MATE estimation error in adversarial cases won't shrink by increasing population size. As a result, the worst-case deviation factor $\xi$ increases with larger population sizes. Therefore, increasing the population size doesn't alleviate the adversarial vulnerability problem. It makes it even worse.  However, increasing the population size is beneficial in another aspect and that is, matching the distributions of covariates in the control and the treatment group becomes more tractable, and higher quality CBMs can be used. It is still worth mentioning that increasing the population size wouldn't alleviate the adversarial vulnerability in any of the given CBMs.

One might naturally think that by introducing randomness, or changing the stop criteria in the CBM procedure, the mentioned adversarial treatment assignments would be less likely to get selected. Examples of this would be to limit the number of iterations in SMD minimization in non-sequential cases, or to select a smaller $p_0$ in Pocock's method. However, it's rather running away from the problem instead of solving it. The gap between the black and blue points in the figure~\ref{fig2} is filled with other possible treatment assignments. Limiting the extent of using CBM would would make it impossible for the current adversarial treatment assignments to be selected, but introduces even worse adversarial assignments. Note that the MATE of black points increases as more imbalance $\mathcal{U}$ is allowed.

\section{Conclusions}

In this work, we have provided arguments to demonstrate that the SMD CBM,and Pocock's sequential assignment method, two of the most used approaches to reduce selection bias in RCTs, are vulnerable to worst-case treatment assignments (Figure~\ref{fig2}). In order to demonstrate these vulnerabilities, we proposed ATASTREET to find well-balanced treatment assignments where the studied CBMs fail in preventing large errors in the MATE. It uncovers a drawback for these CBMs and suggests that these CBMs should not be used to evaluate reliability of the results in RCTs. The worst-caste vulnerability opens up opportunities for deceitful activities to exploit adversarial treatment assignments in order to deviate the measured average treatment effect towards a desired ATE.

We provided an index to check whether a given RCT that used CBM is close to worst-case treatment assignments. We also developed adversarial attacks to any given RCT in order to show that a deceitful researcher can take advantage of worst-case vulnerability.

Our work suggests interesting future research directions. One direction is to evaluate the adversarial robustness of other existing CBMs. Another direction is how to find a CBM with the best adversarial robustness. Such a method is desirable in cases where the nature of the trial has a high importance level that brings the need to use a method which is robust against any deceitful action (e.g. clinical trials in deadly pandemics).

\section{Acknowledgements}
This work was supported by NSF grants 1842378, 1937134, CCF-1911094, IIS-1838177, and IIS-1730574; ONR grants N00014-18-12571, N00014-20-1-2534, and MURI N00014-20-1-2787; AFOSR grant FA9550-18-1-0478, and a Vannevar Bush Faculty Fellowship, ONR grant N00014-18-1-2047.

\bibliography{ref}
\bibliographystyle{ieeetr}

\section{Appendix}
An executive summary of Pocock's sequential treatment assignment method  is given as Algorithm~\ref{pcok}
\subsection{Pocock's sequential assignment method}
\begin{algorithm}[h]
\caption{Pocok's sequential binary treatment assignment. }
\label{pcok}
    \begin{algorithmic}[1]
    \STATE For the few initial subjects, it doesn't matter how to assign them. Randomly assign the few first subjects to the treatment or the control group.
    \STATE A new subject comes to the clinic and the goal is to assign them to one of the groups.
    \STATE Assume the new subject is assigned to either of the groups, e.g the treatment group.
    
     In Pocock's method, covariates are assumed to be categorical. In case some of the covariates are continuous-valued, they should be discretized to different categories.
    \STATE  For each of the covariates, namely the $i^{th}$ covariate, The new subject has the value of the $j^{th}$ category for the $i^{th}$ covariate. Count Number of subjects in the control and in the treatment group having the value of the $j^{th}$ category for the $i^{th}$ covariate. Define $d_i=d(N_{\rm{treatment}},N_{\rm{control}})$ where $d(x,y)$ is a distance function. The most natural case for the binary treatment case is $d(x,y)=|x-y|$.
    \STATE Define $G$ as a (weighted) sum of $d_i$s.   $G=\sum_{i=1}^{m} \alpha_i d_i$. The weights $\alpha_i$ could be arbitrarily selected in order to emphasize balancedness in some of the covariates.
    \STATE Go back to step 3 and this time, assign the new subject to the other group. 
    \STATE Sort two different resulting $G$s for assigning the new subject to each of the groups. Flip a coin with the probability of being head equal to a pre-specified probability of $P_0$. If the coin was head, assign the subject to the group resulting in the smaller $G$;And if it was tail, assign it to the group resulting in the bigger $G$   (Note that $P_0$ should be bigger than 0.5 )
    \STATE For the next subject, go back to step 2 and repeat the same procedure.
    \end{algorithmic}
\end{algorithm}

\subsection{Proof of Theorem1}

\begin{thm}
In Pocock's sequential treatment assignment method, the way a new subject is assigned to a group, minimizes $\mathcal{U}_{\rm{Pocock}}$ with the probability of $p_0$. In other words, $\mathcal{U}_{\rm{Pocock}}$ can be used instead of $G$ in Pocock's method (Algorithm~\ref{pcok}).
\end{thm}

\emph{Proof:}
The goal is to prove $\mathcal{U}_{\rm{Pocock}}$ with
\begin{align}
    \mathcal{U}_{\rm{Pocock}} =\sum_{i=1}^{m} \sum_{j=1}^{N_i} \alpha_i |N_{\rm{control}}^{i}(j) - N_{\rm{treatment}}^{i}(j) |
   \label{eq_Pocock_def} \nonumber
\end{align}
instead of $G$ in Algorithm~\ref{pcok} results in same probability of assigning the subject to each of the treatment or control groups. Assume that the current subject has the value of $c_i$ for the $i^\text{th}$ covariate. Then immediately by the definition of $G$ we have:
\[ G=\sum_{i=1}^{m} \alpha_i d_i = \sum_{i=1}^{m} \alpha_i |N_{\rm{control}}^{i}(c_i) - N_{\rm{treatment}}^{i}(c_i) | \]

By adding and subtracting a term, we can write it as:
\[ =\sum_{i=1}^{m} \sum_{j=1}^{N_i} \alpha_i |N_{\rm{control}}^{i}(j) - N_{\rm{treatment}}^{i}(j) |\]

\[   - \sum_{i=1}^{m} \sum_{j=1, j \neq c_i}^{N_i} \alpha_i |N_{\rm{control}}^{i}(j) - N_{\rm{treatment}}^{i}(j) | \]

\[  = \mathcal{U}_{\rm{Pocock}} - \sum_{i=1}^{m} \sum_{j=1, j \neq c_i}^{N_i} \alpha_i |N_{\rm{control}}^{i}(j) - N_{\rm{treatment}}^{i}(j) |\]

Now note that the second term is a positive number that would remain constant for different assignments of the current subject.
\[ G_2 - G_1 =  \mathcal{U}_{\rm{Pocock},2} - \mathcal{U}_{\rm{Pocock},1}
\]

Thus, $\mathcal{U}_{\rm{Pocock}}$ could be used instead of $G$ in Algorithm~\ref{pcok} and result in the same decision.

We have introduced the adversarial attack to find adversarial treatment assignments in the manuscript, but didn't provide details on how ATASTREET incorporates mixed linear programming to solve the given combinatorial optimization problems. Here, mathematical details for different versions of ATASTREET are provided.

\subsection{ATASTREET for SMD with $\ell_1$}
In order to find adversarial attacks of the SMD with $\ell_1$, one has to solve the optimization problem in Equation~\ref{eq6}

\[ \mathrm{argmax}_{\mathcal{A}} \;  \Bigg( \lambda \; \mathrm{MATE}(  \mathcal{A} ) -  \; \dfrac{2}{N} \left\|   \sum_{ \rm{treatment}}\vec{x}^{\,i} - \sum_{\rm{control}} \vec{x}^{\,i}
    \right\|_{1} \Bigg) 
\]

\[  \mathrm{argmax}_{\mathcal{A}} \;  \Bigg( \lambda \;\sum_{i} (\mathcal{A}_i y_i^1 + (1-\mathcal{A}_i) y_0^1 )  -  \; \left\|   \sum_{i} (2 \mathcal{A}_i -1 )\vec{x}^{\,i} 
    \right\|_{1} \Bigg) 
\]
Then, by throwing away a term that doesn't depend on the $\mathcal{A}$, we can write down the argmax problem as:
\[  \mathrm{argmax}_{\mathcal{A}} \;  \Bigg( \lambda \;\sum_{i} \mathcal{A}_i (y_i^1+y_0^1 )  -  \; \left\|   \sum_{i} (2 \mathcal{A}_i -1 )\vec{x}^{\,i} 
    \right\|_{1} \Bigg) 
\]

By introducing auxiliary variables $t_j^{+},t_j^{-}$,  this argmax problem can then be written as an argmin problem and then be solved using mixed integer linear programming tools.

\[  \mathrm{argmin}_{\mathcal{A},t^{+} , t^{-} } \;  \Bigg( -\lambda \;\sum_{i} \mathcal{A}_i (y_i^1+y_0^1 )  +  \;    \sum_{j=1}^{m} (t_j^{+}+t_j^{-})  \Bigg) 
\]

\[ \forall{j}, \quad  t_j^{+}-t_j^{-}= \sum_{i} (2 \mathcal{A}_i -1 )\vec{x}^{\,i}_{j}
\]
\[  \sum_{i} \mathcal{A}_i = \floor{ \dfrac{N}{2}}
\]
\[ 0 \leq \mathcal{A}_i \leq 1 , \quad 0\leq t_j^{+} , t_j^{-} \; , \quad   \mathcal{A}_i \in \mathbb{N}
\]

\subsection{ATASTREET for SMD with $\ell_\infty$}
In order to find adversarial attacks of the SMD with $\ell_1$, one has to solve the optimization problem in Equation~\ref{eq6}

\[ \mathrm{argmax}_{\mathcal{A}} \;  \Bigg( \lambda \; \mathrm{MATE}(  \mathcal{A} ) -  \; \dfrac{2}{N} \left\|   \sum_{ \rm{treatment}}\vec{x}^{\,i} - \sum_{\rm{control}} \vec{x}^{\,i}
    \right\|_{\infty} \Bigg) 
\]

\[  \mathrm{argmax}_{\mathcal{A}} \;  \Bigg( \lambda \;\sum_{i} (\mathcal{A}_i y_i^1 + (1-\mathcal{A}_i) y_0^1 )  -  \; \left\|   \sum_{i} (2 \mathcal{A}_i -1 )\vec{x}^{\,i} 
    \right\|_{\infty} \Bigg) 
\]
Then, by throwing away a term that doesn't depend on the $\mathcal{A}$, we can write down the argmax problem as:
\[  \mathrm{argmax}_{\mathcal{A}} \;  \Bigg( \lambda \;\sum_{i} \mathcal{A}_i (y_i^1+y_0^1 )  -  \; \left\|   \sum_{i} (2 \mathcal{A}_i -1 )\vec{x}^{\,i} 
    \right\|_{\infty} \Bigg) 
\]

By introducing auxiliary variables $t_j^{+},t_j^{-} , \,T$,  this argmax problem can then be written as an argmin problem and then be solved using mixed integer linear programming tools.

\[  \mathrm{argmin}_{\mathcal{A} , T , t_j^{+} , t_j^{-}} \;  \Bigg( -\lambda \;\sum_{i} \mathcal{A}_i (y_i^1+y_0^1 )  +  \; T  \Bigg) 
\]
\[ \forall{j}, \quad  t_j^{+}-t_j^{-}= \sum_{i} (2 \mathcal{A}_i -1 )\vec{x}^{\,i}_{j}
\]
\[  \sum_{i} \mathcal{A}_i = \floor{ \dfrac{N}{2}}
\]
\[ \forall{j}, \quad t_j^{+}+t_j^{-} \leq T
\]
\[ 0 \leq \mathcal{A}_i \leq 1 , \quad 0\leq t_j^{+} , t_j^{-} , T\; , \quad   \mathcal{A}_i \in \mathbb{N}
\]

\subsection{ ATASTREET for $\mathcal{U}_{\rm{Pocock}}$}
In order to find adversarial attacks of the Pocock's assignment method, one has to solve the optimization problem in Equation~\ref{eq_atastreet_pocock}

\[ \mathrm{argmax}_{\mathcal{A}} \;  \Bigg( \lambda \; \mathrm{MATE}(  \mathcal{A} ) -  \; \sum_{i=1}^{m} \sum_{j=1}^{N_i} \alpha_i |N_{\rm{control}}^{i}(j) - N_{\rm{treatment}}^{i}(j) | \Bigg) 
\]

In order to implement $\mathcal{U}_{\rm{Pocock}}$ in a linear format, we write it as:
\[ \mathcal{U}_{\rm{Pocock}}= \left\| \tilde{X} (2\vec{\mathcal{A}}-1)\right\|_1
\]
Where $\tilde{X}$ is a matrix formed as below
\begin{align}
    \tilde{X}=
    \begin{bmatrix}
        d_1^1(1) & d_1^1(2) & \dots & d_1^1(N)\\
        d_2^1(1) & d_2^1(2) & \dots & d_2^1(N)\\
        \vdots   &          &\ddots & \vdots \\
        d_{N_1}^1(1) & d_{N_1}^1(2) & \dots & d_{N_1}^1(N)\\
        \hline\\
        d_1^2(1) & d_1^2(2) & \dots & d_1^2(N)\\
        d_2^2(1) & d_2^2(2) & \dots & d_2^2(N)\\
        \vdots   &          &\ddots & \vdots \\
        d_{N_2}^2(1) & d_{N_2}^2(2) & \dots & d_{N_2}^2(N)\\
        \hline\\
        \vdots   &          &\ddots & \vdots \\
        \hline\\
        d_1^m(1) & d_1^m(2) & \dots & d_1^m(N)\\
        d_2^m(1) & d_2^m(2) & \dots & d_2^m(N)\\
        \vdots   &          &\ddots & \vdots \\
        d_{N_m}^m(1) & d_{N_m}^m(2) & \dots & d_{N_m}^m(N)
    \end{bmatrix}
   \label{eq_Amat}
\end{align}

where $d_j^i(k)= 1 \iff$ $k^\textit{th}$ subject has the $j^\textit{th}$ category for  $i^\textit{th}$covariate.

Similar to previous subsection, by introducing auxiliary variables $t_j^{+},t_j^{-}$,  this argmax problem can then be written as an argmin problem and then be solved using mixed integer linear programming tools.

\[  \mathrm{argmin}_{\mathcal{A} , t_j^{+} , t_j^{-}} \;  \Bigg( -\lambda \;\sum_{i} \mathcal{A}_i (y_i^1+y_0^1 )  +  \; \sum_{j=1}^{N_0+\dots+N_m} (t_j^{+} + t_j^{-}) \Bigg) 
\]
\[ 1\leq \forall{j}, \leq N_0+\dots+N_m \quad  t_j^{+}-t_j^{-}= \sum_{i} (2 \mathcal{A}_i -1 )\tilde{X}(j,i)
\]
\[  \sum_{i} \mathcal{A}_i = \floor{ \dfrac{N}{2}}
\]
\[ 0 \leq \mathcal{A}_i \leq 1 , \quad 0\leq t_j^{+} , t_j^{-} \; , \quad   \mathcal{A}_i \in \mathbb{N}
\]

\end{document}

%% file: math.tex
\usepackage{amsmath,amsfonts,bm, bbm}

\def\eqref#1{equation~\ref{#1}}

\def\floor#1{\lfloor #1 \rfloor}
\def\1{\bm{1}}

\DeclareMathAlphabet{\mathsfit}{\encodingdefault}{\sfdefault}{m}{sl}
\SetMathAlphabet{\mathsfit}{bold}{\encodingdefault}{\sfdefault}{bx}{n}

\def\E{{\mathbb{E}}}

\usepackage{mdframed}

\newtheorem{thm}{Theorem}

\usepackage{accents}
\newlength{\dhatheight}